# Customer Churn in World of Warcraft

by Sulman Khan

3/31/2020

## Abstract

World of Warcraft is a massively multiplayer online video game released on November 23, 2004 by Blizzard Entertainment. In contrast with traditional games only having a single upfront fee to play, WoW also has a monthly subscription to play the game. With customer subscriptions in mind, we can apply the use of churn prediction to not only predict whether a customer will unsubscribe from the service but explore the user's playing behavior to obtain more insight into user playing patterns. The churn problem is somewhat complex due to the nature of not having a one size fits all solution – as different services define churn in a variety of ways. In this paper, we explore a dataset that focuses on one year from January 1, 2008 until December 31, 2008, as it highlights the release of a major content update in the game. Machine learning is used in two aspects of this paper: Survival Analysis and Binary Classification. Firstly, we explore the dataset using the Kaplan Meier estimator to predict the duration until a customer churns, and lastly predict whether a person will churn in six months using traditional machine learning algorithms such as Logistic Regression, Support Vector Machine, KNN Classifier, and Random Forests. From the survival analysis results, WoW customers have a relatively long duration until churn, which solidifies the addictiveness of the game. The best binary classification algorithm achieved a 96% ROC AUC score in predicting whether a customer will churn in six months.

# Introduction

World of Warcraft is a massively multiplayer online video game released on November 23, 2004. Before this era, MMORPG's catered to a small segment of video gamers. But with the massive success of WoW, various video game companies decided to invest resources into developing large-scale titles. Video games were sought out as movie-like experiences, where you follow a single protagonist. However, WoW did not follow a single protagonist, but all the users playing the video game. Not only was the main objective different from single-player games, but the pricing model. Traditional games followed a single upfront fee. In addition to the single upfront fee, WoW had a monthly subscription to play the game. The monthly subscription was warranted because of the constant updates to the game and server fees.

Now, how can we use machine learning to provide actionable business insight, such areas are presented below.

(1) Player Segmentation
(2) Server Optimization
(3) Churn Prediction

Player segmentation uses clustering methods to segment player behavior. Player behavior provides insight into what areas, the game developers should spend resources updating.

In beginning your journey in WoW, players are tasked with selecting their realm and will only be playing with the same players in that realm. At the time, this was an issue with server infrastructure as you do not want to overload these physicals servers with 3000+ players. Server optimization is a problem in properly utilizing the server infrastructure to optimize the latency or delays players receive in-game. A solution to this problem is to provide a separate server for regions with high player densities to provide less latency for players. But additional servers for each zone could be costly, but with the use of machine learning we can note highly congested areas at various times in the day and implement a server to handle the load for a brief period, it occurs.

Lastly, churn predictions relate to customer retention, specifically predicting if a customer will leave the service. The churn problem is somewhat complex due to the nature of not having a one size fits all solution – as different services define churn in a variety of ways. Knowing customer churn can provide significant insight into grouping certain behaviors in the video game to prevent it. In this paper, we will explore churn concerning survival analysis and binary classification.

# Dataset

The data originates from a Taiwanese realm on the Horde faction over a three-year period. Each data timestamp was collected by using the in-game querying tool. Every 10 minutes, the data was queried resulting in 144 timestamps per day. Limitations of the query tool include only listing 50 entries, thus requiring additional querying and filtering to acquire a total view. The initial dataset contains several subdirectories containing logs for each date, which requires additional preprocessing for exploratory data analysis. The WoWAH-parse.py script was used to load all the traces in the avatar history subdirectories and merge them [1]. We will focus on the one-year period from January 1, 2008 until December 31, 2008, as it highlights a player population peak with the launch of a major content expansion, Wrath of the Lich King.

The dataset is comprised of multiple features such as:

- char – unique character identifier
- level – current level of the character
- race – character's selected race
- charclass – character's class
- zone – current location in which the character resides
- guild – unique guild identifier
- timestamp – date and time when the data was queried

The number of unique entries for each feature is organized in Table 1.

Table *1*: Feature descriptions.

| Unique Characters | Unique Levels | Unique Races | Unique Classes | Unique Zones | Unique Guilds | Unique Timestamps |
|---|---|---|---|---|---|---|
| 37354 | 80 | 5 | 10 | 158 | 420 | 1826400 |

Before exploratory data analysis, the timestamps feature was converted into a DateTime object, allowing us to extract the month, day, and time of each entry. Additionally, the leveling feature was discretized into leveling intervals for each character.

# Exploratory Data Analysis

The EDA section provides reasonable insight into game balance and feature exploration for the predictions section.

## Leveling Distribution

The level feature was explored with a distribution plot in Figure 1. Most players are level 1 with 11598 out of 37354 character entries because users create alternate characters, which are used as mules in major towns to sell and trade items between others. Also, with such a right-skewed distribution it may be informative for game developers to put more effort into the level 1 – 30 zones. There appears to be a steady decrease from 1 to 60 until 70, which was the max level before the WOTLK expansion. Finally, a peak at 80, which is the latest max level.

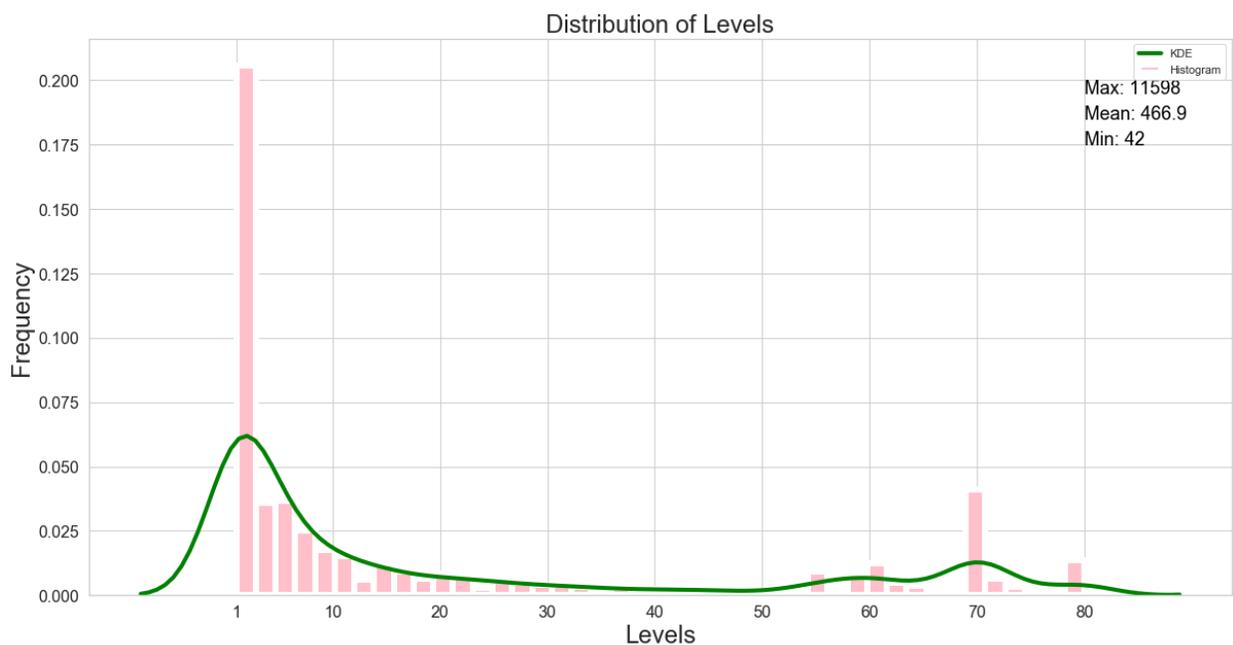

Figure 1: The leveling distribution of players.

The frequency of level intervals is presented in Figure 2. The zones which occupy WoW are not for individual levels but comprises level intervals. From observing Figure 2, the top three level intervals are 0-9, 10 – 19, and 70-79. Such that having insight into how the population is divided in these intervals, further supports our initial reasoning to provide more development time in these early-level zones.

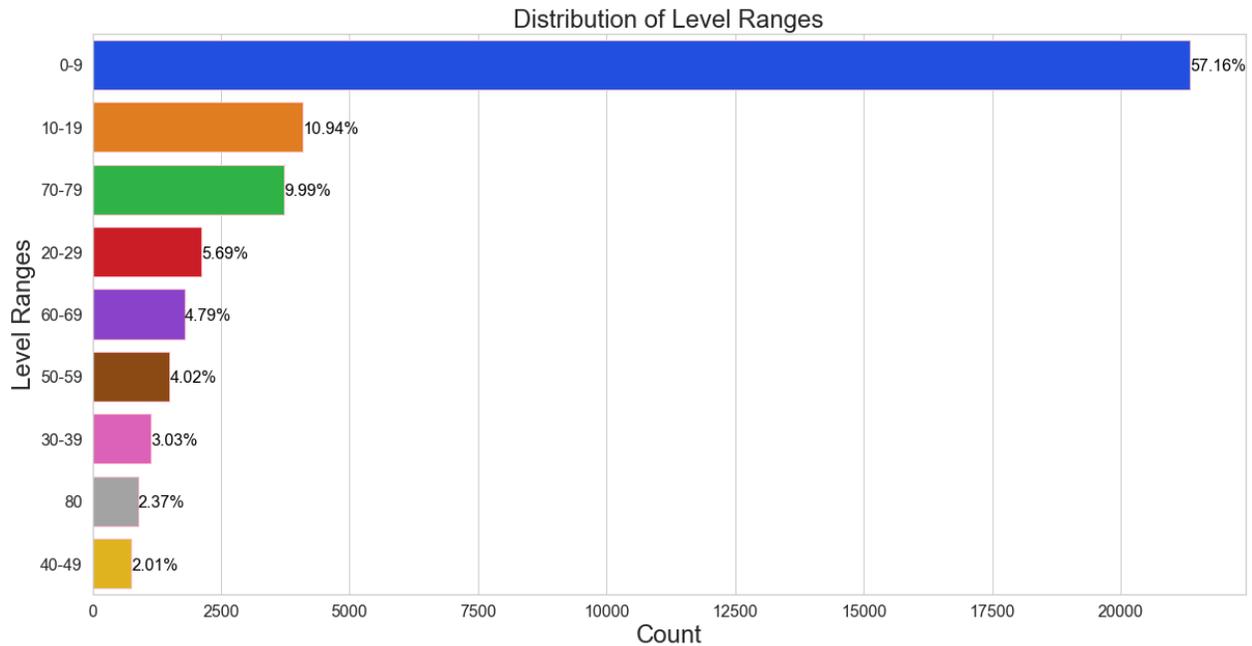

Figure 2: The distribution of level intervals.

*Character Specific Distributions*

Multiple character specific distributions regarding classes, races, and character + race combinations are presented in Figure 3 (a,b,c). When players create their characters, they have a choice in selecting the race and class. Having data on their selections provides developers with insight to properly manage what classes should be monitored more and helps the game balance team to decide on why these certain classes or races are picked – Are they strong, weak, or people enjoy the themes of the classes. Should we spend more time on their race-specific starting zones due to their selection? From Figures 3(a,b,c) the player base selects Blood Elves, Warriors, and Orc Warriors, as their popular choices in races, classes, and race-class combinations, respectively.

(a) Races

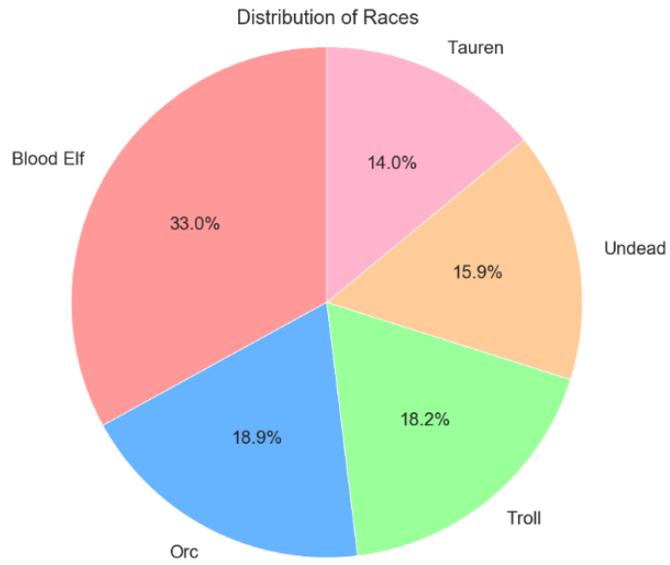

(b) Classes

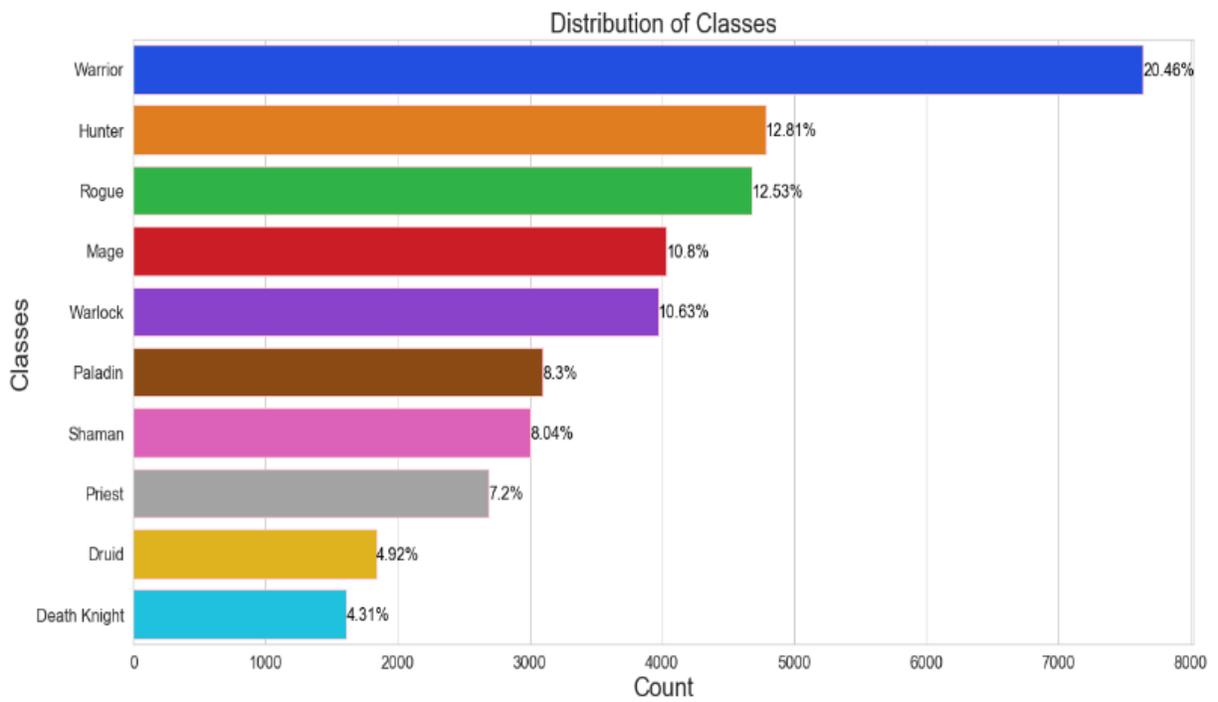

(c) Race-Class Combinations

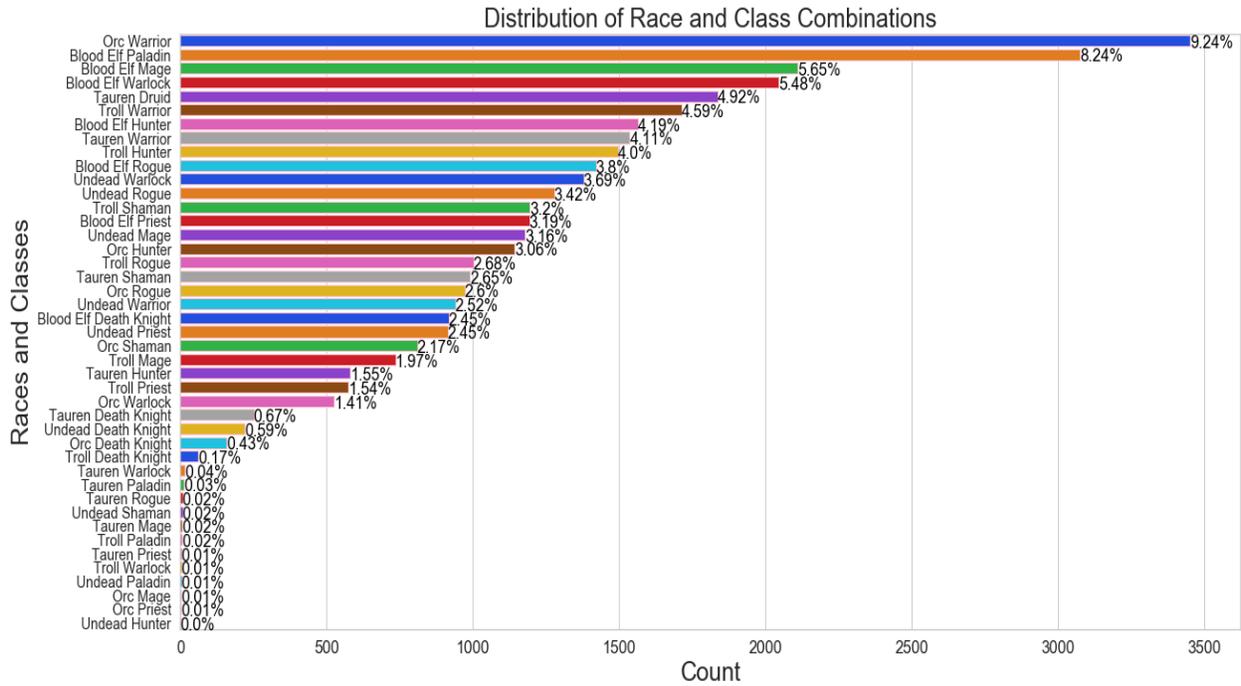

Figure 3: The distributions of (a) races, (b) classes, and (c) race-class combinations.

The population of zones is investigated in Figure 4. Identifying population bottlenecks is important for optimizing server stability, but investigating which zones are populated is crucial in delegating resources into what aspects of the game players are enjoying. The game is divided into Player vs Environment and Player vs Player content. The top two zones are hubs where players can trade items or sign up for quests. The third is a major PvE raid dungeon where players embark on a quest to clear the opposing forces.

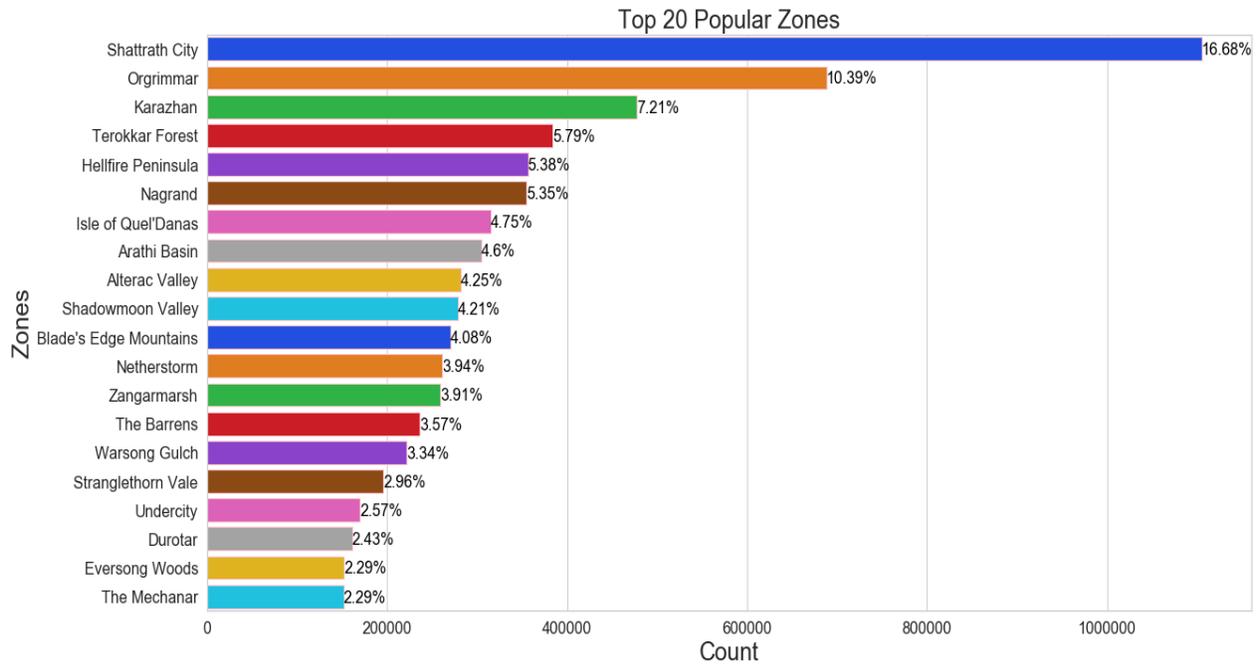

Figure 4: The distribution of zones.

*Guild Specific Distributions*

The guild population and class distributions in guilds are presented in Figure 5 (a,b). There are roughly 500 guilds created - The largest guild has a population of 1796 members, and the average guild has 73 members. Similar to the class distribution in Figure 3 (b), guilds follow the same breakdown. Higher-level content requires an abundant amount of people requiring 25 or more and use guilds to gather people. Not only are guilds used for completing WoW activities, but many users have it function as a social community to keep in touch with their online friends.

(a): Guild Population

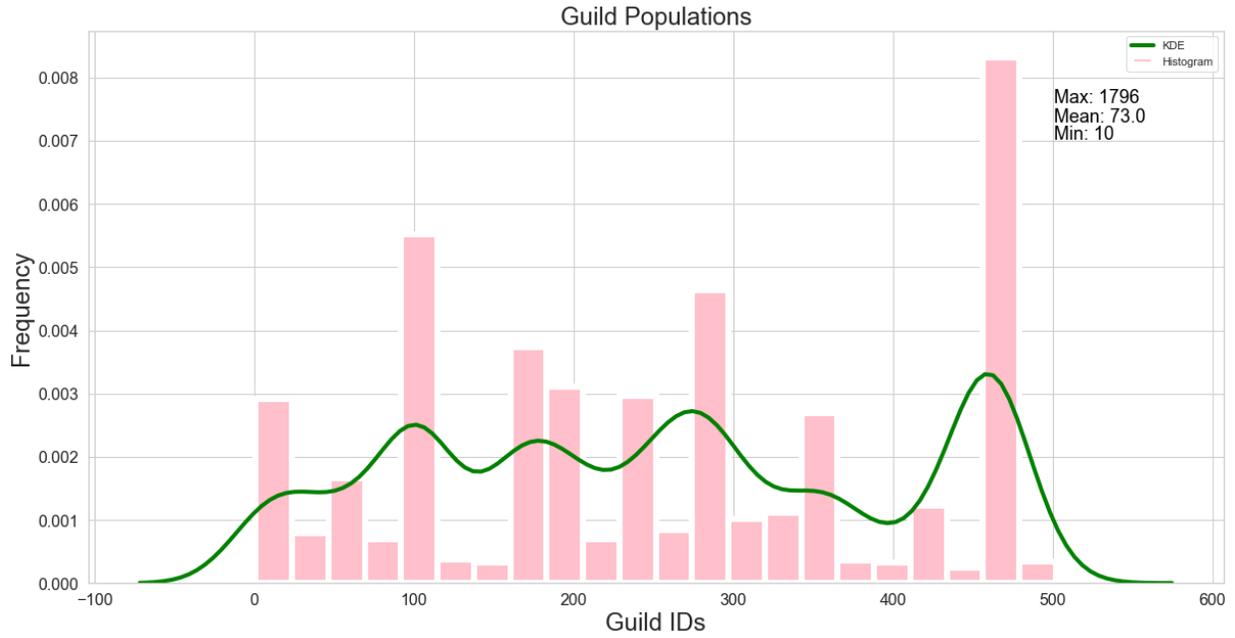

(b) Class distribution in guilds

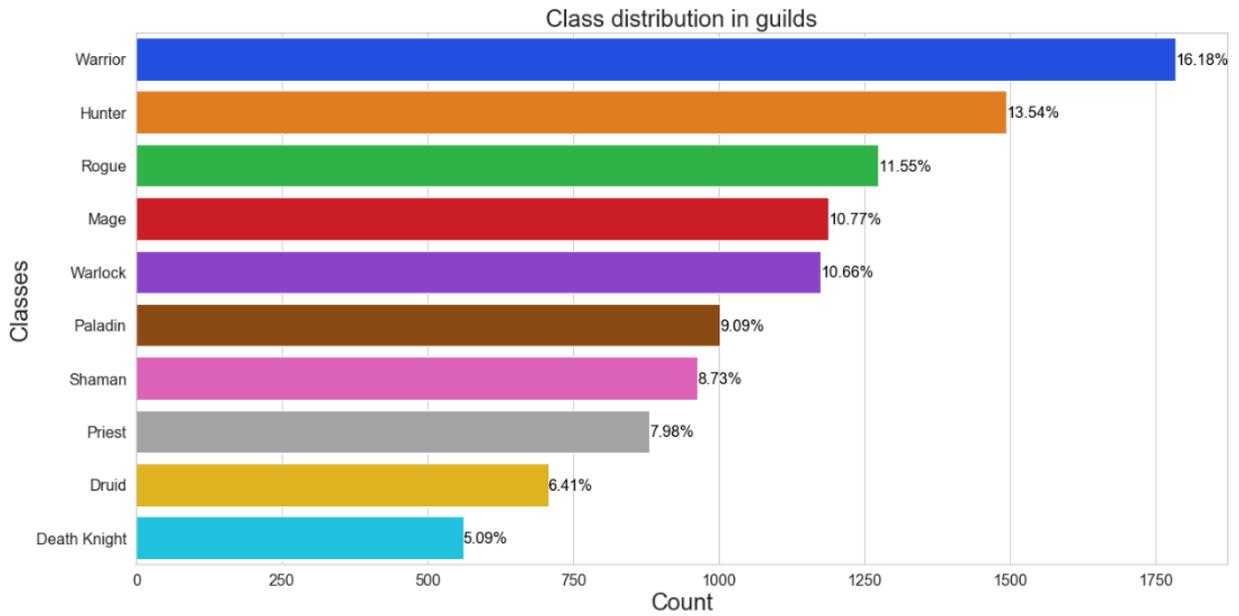

Figure 5: (a) frequency of users in guilds and (b) class distributions in guilds.

*Player Activity*

The daily and monthly activity of players in their respective intervals are presented in Figure 6 (a,b). There is a steady equilibrium of players between 70-79 having the highest frequency because the max level before November was 70. The max level was increased to 80 in November. Similar fluctuations are observed within the other leveling intervals between January and September, but a sharp peak in the 0-9 interval arises during October. A possible reason is the arrival of new players who are awaiting the expansion release in November.

(a) DAU

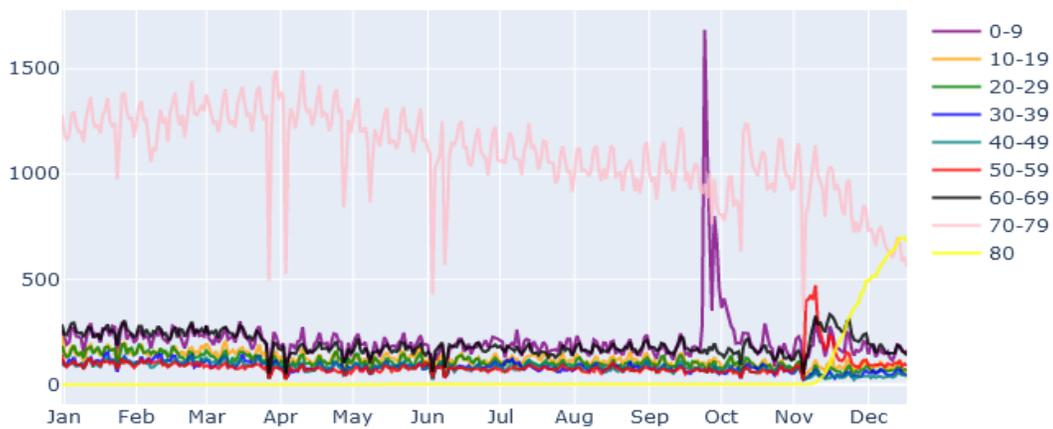

(b) MAU

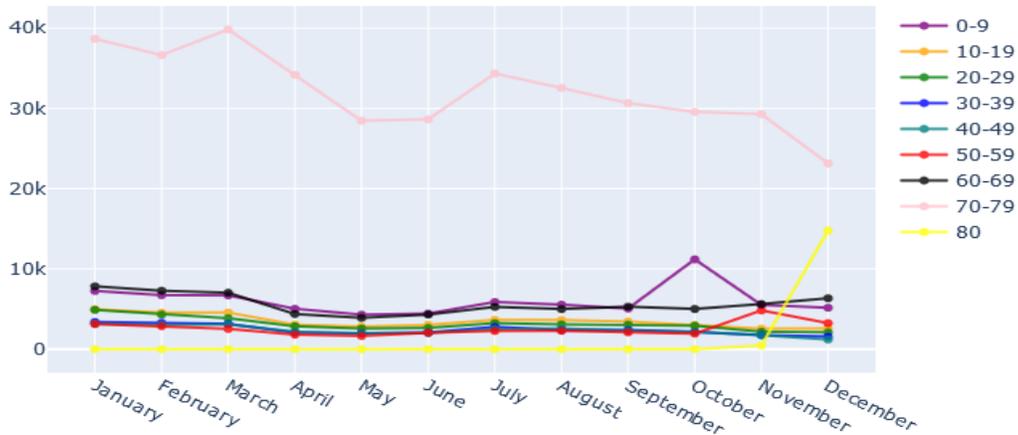

Figure 6: (a) daily user activity and (b) monthly user activity over a one-year period.

The hourly user activity, daily playtime of users, and weekday frequency of users are presented in Figure 7(a,b,c). From observing the HAU, there is a steady decrease until 7:00 am and then elevates during the afternoon because teenagers arrive from school. Then, a steady increase in the evening from 5:00 pm until 12:00 pm because adults arrive from work. The highest frequency of users is during the evening hours. The average user plays for 1 hour daily, and the 95% percentile of players play for 4 hours daily. Having this knowledge allows us to properly address times in which the servers need to be taken down for maintenance, and how to pace future content updates for the average user's playtime - do not want to release content which the average user cannot finish. Most users are recorded playing on Sunday, and the other days of the week are in similar in frequency. We'd expect most players to play on the weekend to account for not having work obligations.

(a) HAU

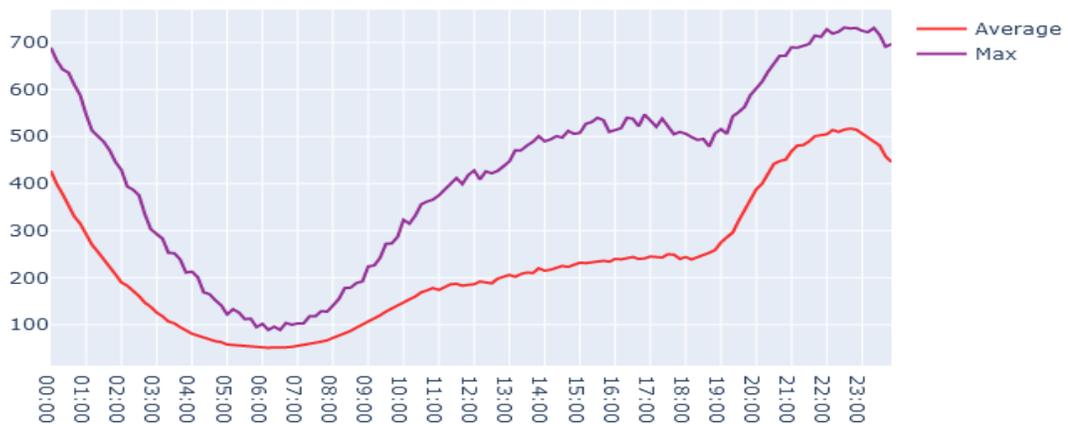

(b) Daily Playtime

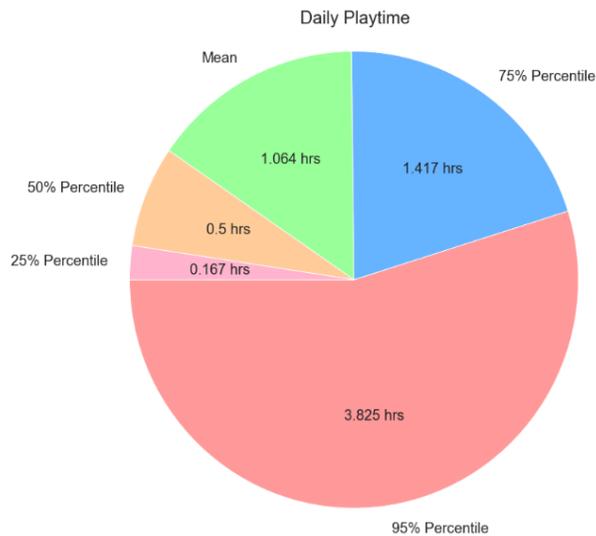

(c) User Frequency on Weekdays

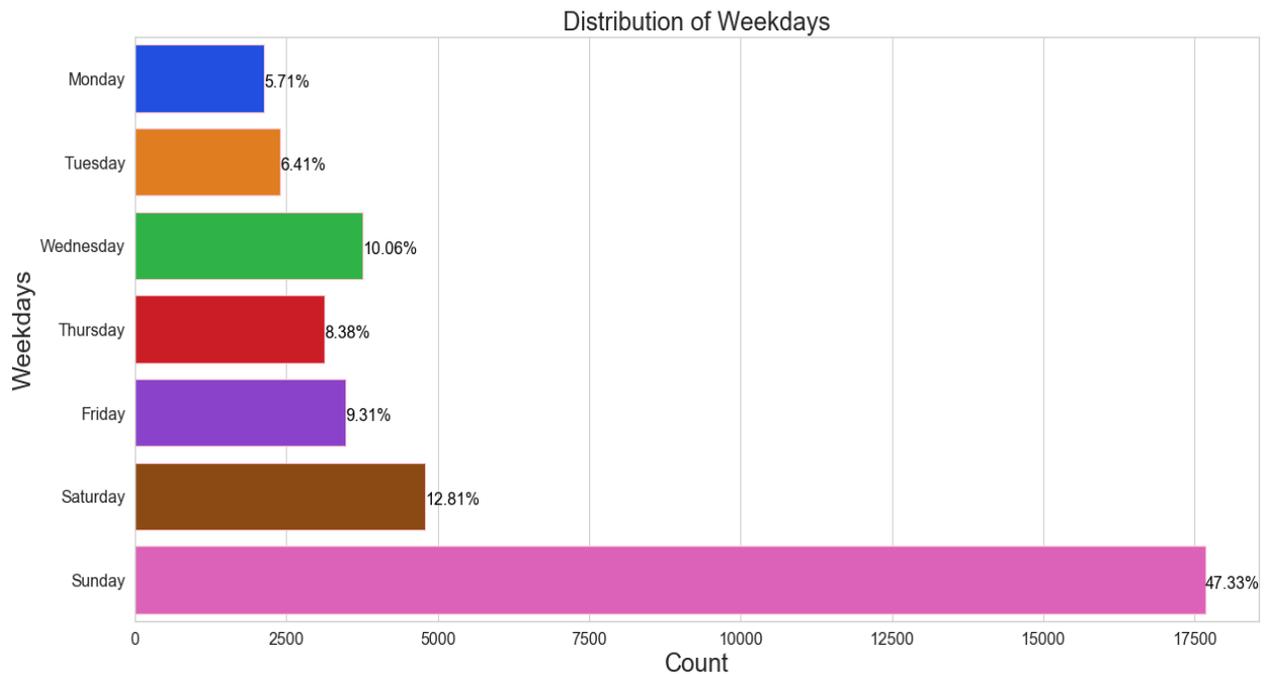

Figure 7: (a) hourly user activity, (b) daily playtime, and (c) user frequency during days of the week.

## *Survival Analysis*

The main objective of survival analysis is to provide an accurate experimental analysis when the collected data has not been completed. It was inspired for applications concerning clinical trials in the diagnosis of illnesses, and how to properly classify patients if they do not exhibit symptoms during the lifetime of the study. This is known as censoring, and there are three ways to define this issue [2].

1. the observed event does not occur for an object till the end of a study
2. an object was eliminated from study or decided to abandon the study
3. an object is lost from observation before the end of the study (eg. he died for the reason other than that of the event under study)

In completion of such experiment results in data:

1. survival time (means a period of individual observation; also used for censored data)
2. The reason why the individual observation was terminated – coded as 1 for the observed event, 0 for patients from groups 2 and 3.

Using the above clauses allows us to construct a survival function, which predicts the probability a person will unsubscribe in our case. The Kaplan-Meier estimator will be used to predict the survival function, which assumes that the pattern of censoring is independent of the survival times. The equation for the Kaplan-Meier estimator is presented in Equation 1 [2]. Successive probabilities are multiplied with previously computed probabilities to get the final estimate.

$$s_t = \frac{N_{\text{subjects, alive}} - N_{\text{subjects, dead}}}{N_{\text{subjects, alive}}} \qquad \text{Equation (1)}$$

There are two verified approaches in comparing KM curves:

- Log-rank test
- Restricted Mean Survival Times

The log-rank test statistic compares estimates of the hazard functions of the two groups at each observed event time. It is constructed by computing the observed and expected number of events in one of the groups at each observed event time and then adding these to obtain an overall summary across all-time points where there is an event. However, the log-rank test statistics fails when the survival curves overlap.

The restricted mean is a measure of average survival from time 0 to a specified time point and may be estimated as the area under the survival curve up to that point [3]. In comparison of two survival curves, we can take the ratios of the RMST to predict the churn ratio in our case.

Before KM survival analysis, the data was filtered discard users who are on a trial period and have not subscribed yet (< 30 days). Also, various churn periods were assigned to a person if they did not play in 2, 3, 4, or 6-month periods.

### *Results*

The KM survival curves with various churn periods are presented in Figure 8. The data appears to be heavily right-censored and does converge to a constant instead of reaching to 0. The 2-month and 3-month churn periods provide more accurate results due to less censoring than the other survival curves. The 2-month churn period has a 70% probability of not churning for ~215 days, and the 3-month period has a 60% probability of not churning for ~245 days. These results display promising insights into how addictive this game is as the population will be healthy across ~200 days.

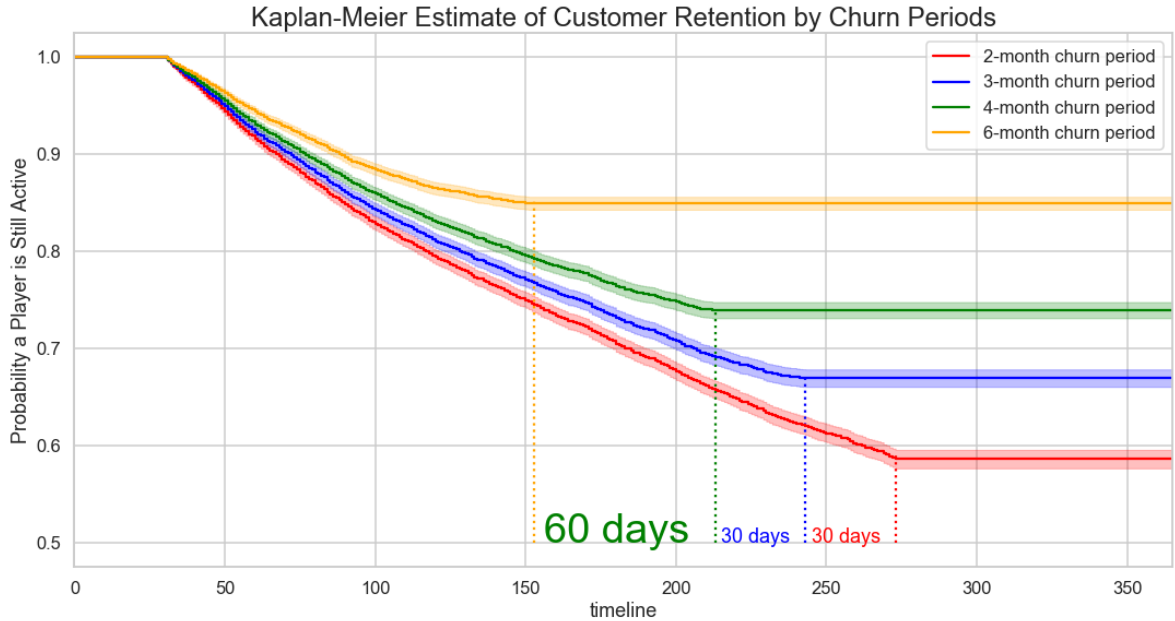

Figure 8: KM survival curves with various churn periods.

The survival analysis curve is plotted for the guild label in Figure 9. People who are not in a guild have a 1.33x higher churn ratio because of not being in a social network – can't interact and group with your guildmates when playing the game.

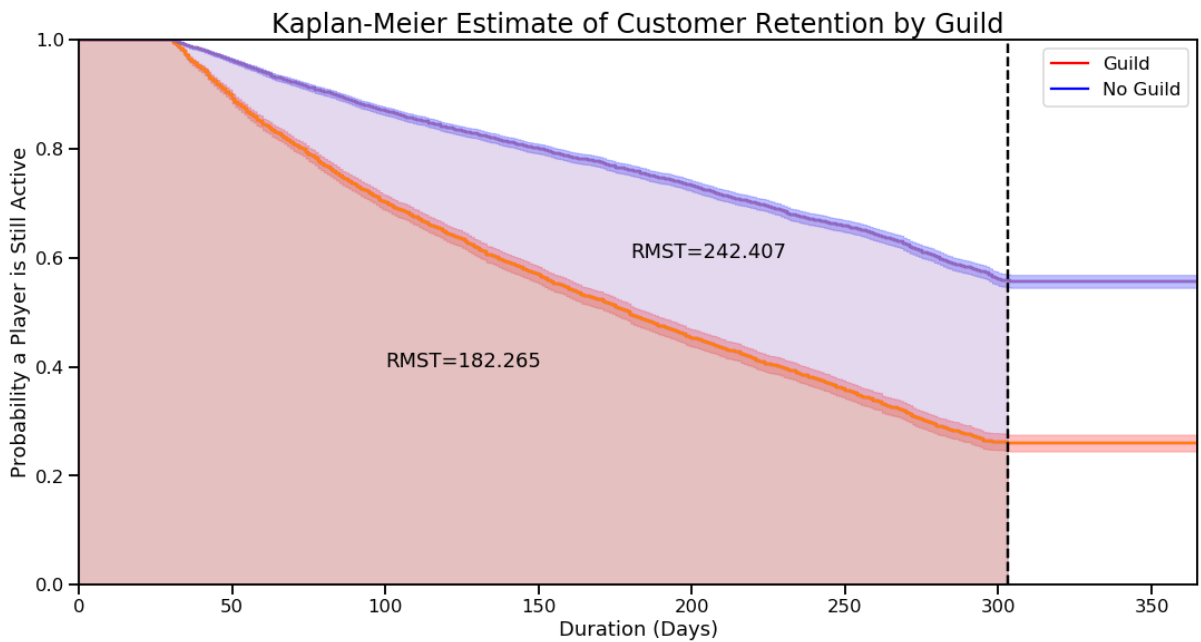

Figure 9: KM survival curves with users not in a guild and users in a guild.

The survival analysis curves are plotted for various level intervals in Figure 10 (a). There is a significant overlap between curves. Therefore, let's compare each curve with the 70-79 range to compare the churn ratio in Figure 10 (b). The churn ratios are listed in Table 2. An interesting detail is the survival analysis curve for the 10-19 interval has a higher churn ratio than the other intervals. A possible solution is updating the 10-19 range content to compensate for this higher churn ratio.

(a) KM survival curves with various level intervals.

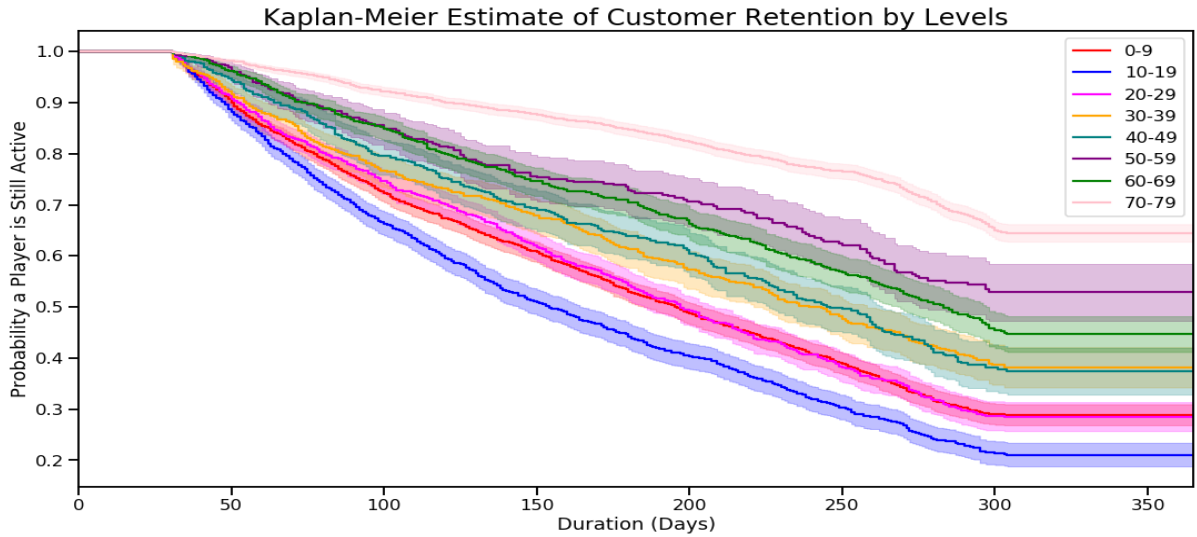

(b) KM survival curves compared with the 70–79 level interval.

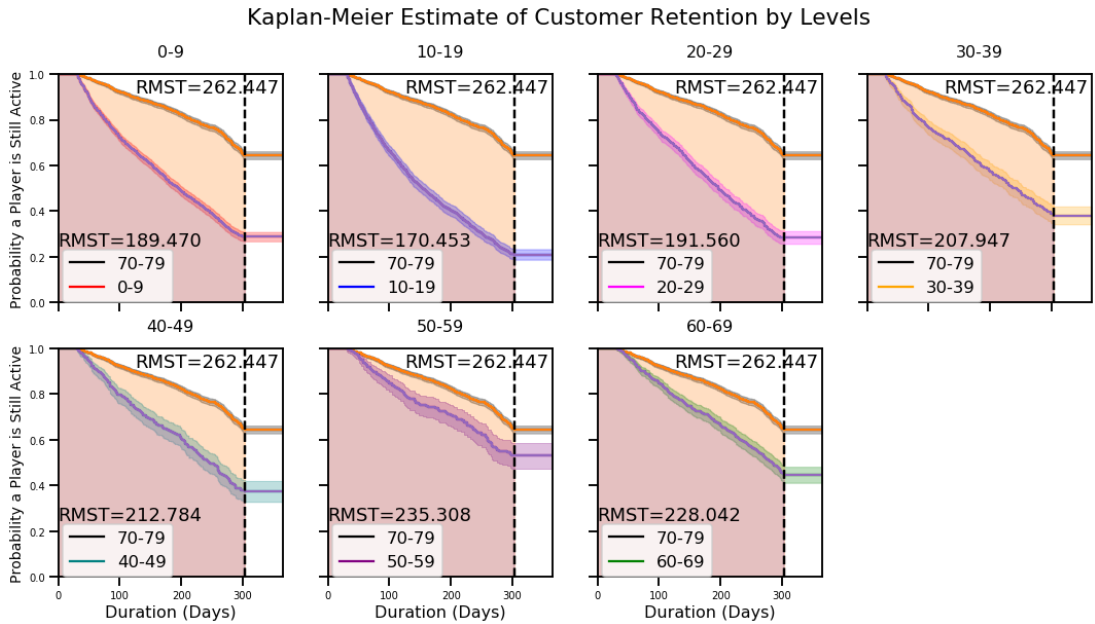

Figure 10: (a) KM survival curves with various level intervals. (b) KM survival curves compared with the 70–79 level interval.

Table 2: Churn ratios with respect to various level intervals.

| Level Interval | Churn Ratio |
|---|---|
| 0-9 | 1.385 |
| 10-19 | 1.54 |
| 20-29 | 1.37 |
| 30-39 | 1.262 |
| 40-49 | 1.233 |
| 50-59 | 1.115 |
| 60-69 | 1.151 |

The survival analysis curves for average playing hours and average playing densities are presented in Figure 11 (a,b). From Figure 11 (a), the churn ratios are 1.076x, and 1.1713x with users who have more than two average hours played. From Figure 11 (b), the churn ratio is 1.2723x with users who have greater than or equal to a 0.5 average playing density. An explanation for this observed behavior is players with more average hours played or average playing densities would exhibit lower churn ratios because they are spending more time playing the game and less likely to unsubscribe.

(a) KM survival curves with various average hours played.

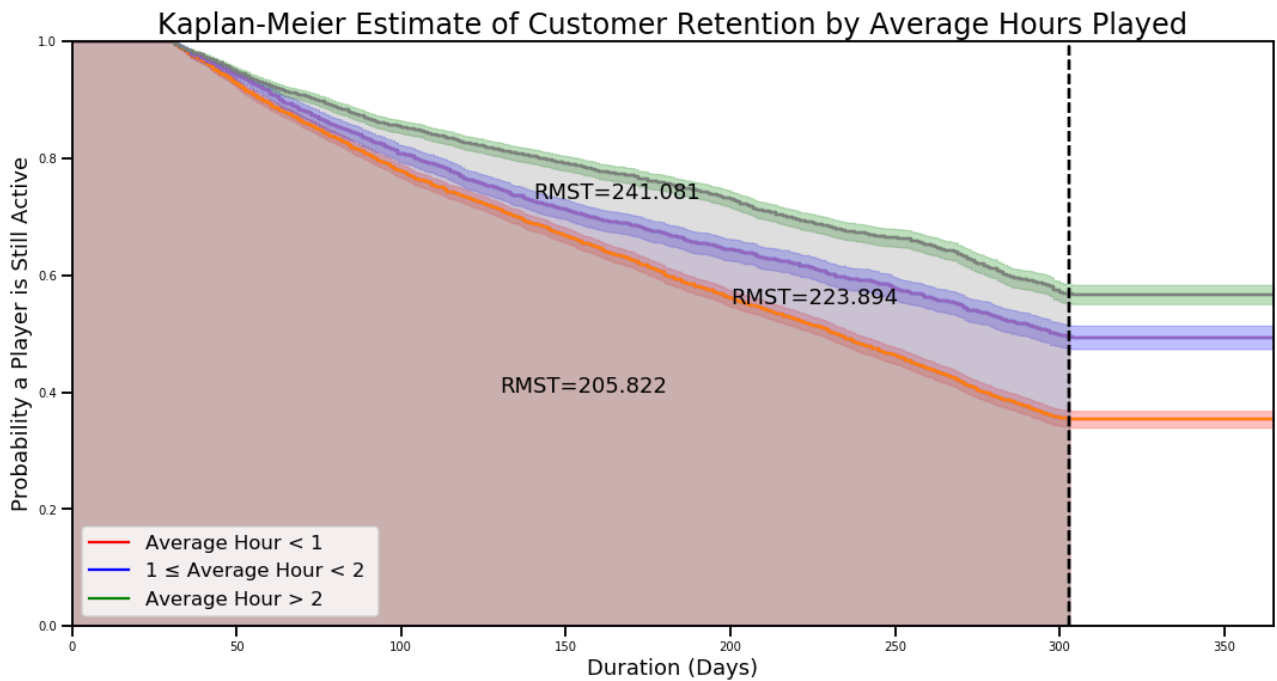

(b) KM survival curves with various average playing densities.

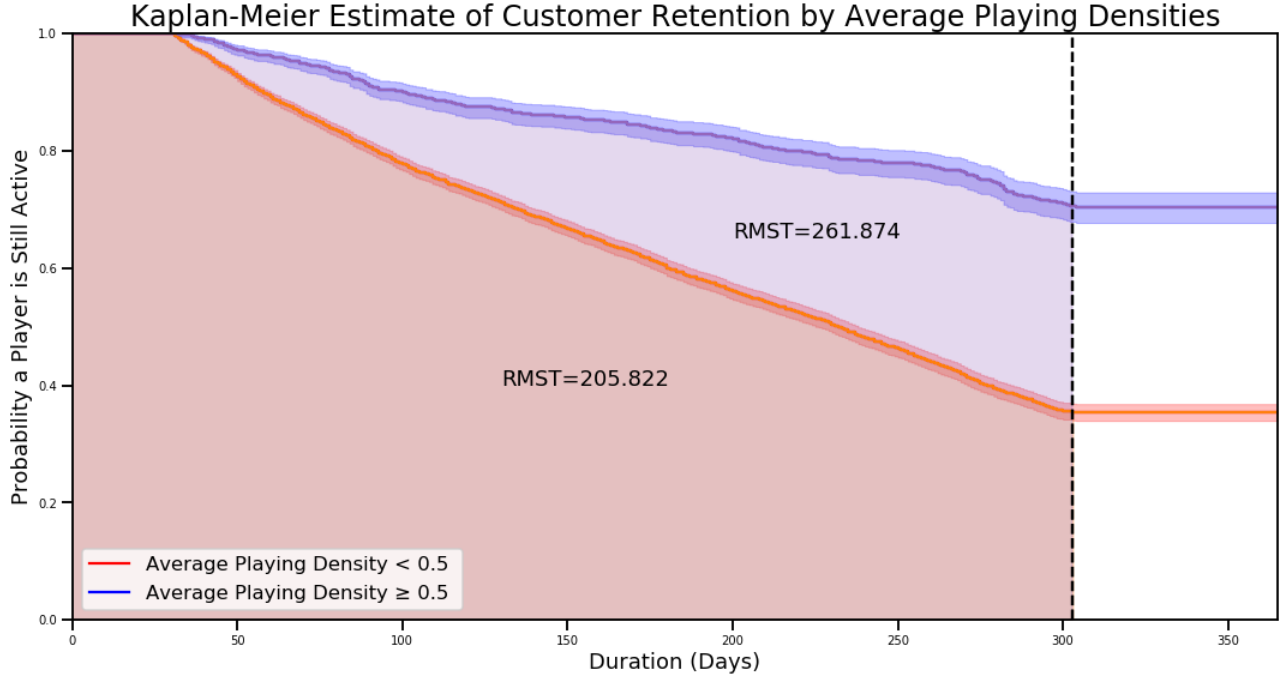

Figure 11: (a) KM survival curve with various average hours played. (b) KM survival curves with various average playing densities.

## Churn Prediction

*Binary Classification*

Binary classification is a problem involving the assignment of a sample to one of two categories by measuring a series of attributes [4]. The semantics include learning a function that minimizes the misclassification probability of labeling a sample as a 0 or 1. The main metric selected is the Receiver Operating Characteristic (ROC) curve, which plots the true positive rate and false-positive rate. The true positive rate is defined as the recall and is presented in Equation 2 [4]. The false-positive rate is presented in Equation 3. The ROC curve plots the TPR vs FPR by varying classification thresholds – lowering the classification threshold results in a higher positive rate. Using the ROC curve, we'll measure the area under the curve (AUC) because it provides an aggregate measure of performance across all possible classification thresholds. Two reasons for using AUC:

1. AUC is scale-invariant - it measures how well predictions are ranked.
2. AUC is classification-threshold-invariant - it measures the quality of the model's predictions without selecting a classification threshold.

In our problem, misclassification of false positives (labeling someone who has quit the game as playing) is more important because it affects the projected finances of the game as the model is predicting a user is playing but is not.

$$\text{TPR} = \frac{\text{TP}}{\text{TP} + \text{FN}} \qquad \text{Equation (2)}$$

$$\text{FPR} = \frac{\text{FP}}{\text{FP} + \text{TN}} \qquad \text{Equation (3)}$$

*Sample Split*

The samples are split into an 80/20 ratio between the training and test samples, respectively. There was k-fold cross-validation performed on the training set for hyperparameter tuning. In k-fold cross-validation, the data is evenly split into k non-overlapping pieces and each piece contains a validation set as displayed in Figure 12 [5]. Then, average the selected scoring metric across the k trials. Cross-validation was used to test the generalizability of the model. As CV checks in how it is performing on new unseen data with a limited amount of training samples. Since the problem is binary classification, stratification can be used to make sure the same number of proportion of classes are in the validation and training folds.

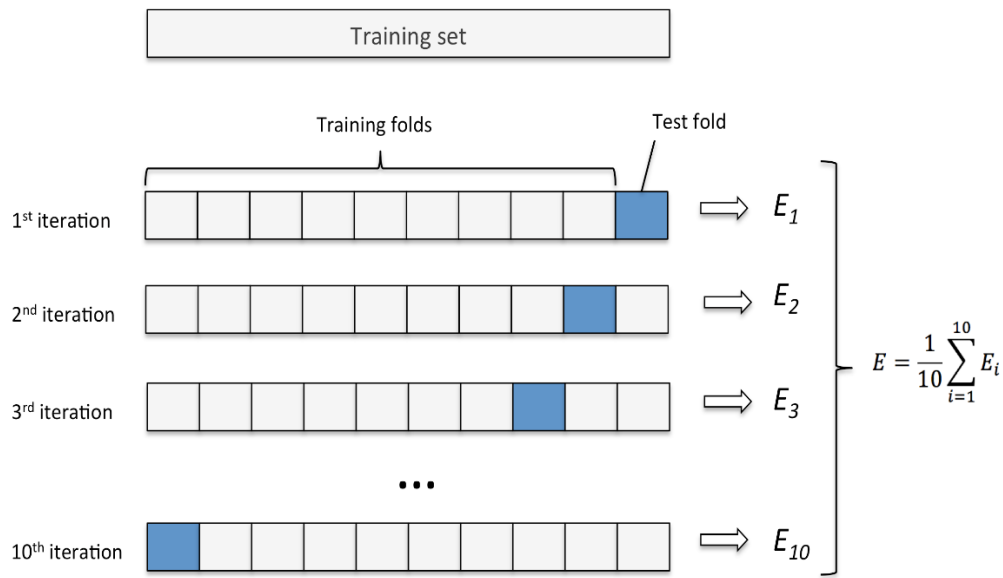

Figure 12: Schematic of 10 – fold cross validation.

*Feature Selection*

Two techniques were compared for feature selection: univariate feature selection and recursive feature elimination. Univariate feature selection works best when you have an abundant number of features and want to select which features are significant using a statistical test. Recursive feature elimination is utilized for identifying which features are increasing your performance metric. Recursive feature selection works by eliminating the least important features. It continues recursively until the specified number of features is reached.

*Machine Learning Models*

<u>Logistic Regression</u>

Having no relationship to its name Logistic Regression is a method used for classification. The output of a logistic function is between 0 and 1 and is presented in Equation 2 [6].

$$\text{logistic}(\eta) = \frac{1}{1 + exp(-\eta)} \qquad \text{Equation (4)}$$

The weighted sum is transformed by the logistic function to a probability. Instead of using the difference in two predictions, the ratio is taken and is displayed in Equation 3.

$$\frac{odds_{x_j+1}}{odds} = \frac{exp\left(\beta_0 + \beta_1 x_1 + \ldots + \beta_j(x_j + 1) + \ldots + \beta_p x_p\right)}{exp\left(\beta_0 + \beta_1 x_1 + \ldots + \beta_j x_j + \ldots + \beta_p x_p\right)} \qquad \text{Equation (5)}$$

*Hyperparameters*

- The penalty is the penalty function used for regularization. The two penalty functions are L1 and L2 and represent the absolute value of the magnitude and squared magnitude loss functions, respectively. The difference between these two techniques is that L2 shrinks the less important feature's coefficient to zero, thus, removing some features altogether.
- Regularization is the coefficient used in the penalty function.

The hyperparameters were tuned to find the best ROC AUC score for logistic regression using an exhaustive grid search (10-fold cv). The optimal hyperparameters are listed below.

- Penalty: L1
- Regularization: C = 25

<u>Support Vector Machine</u>

Support Vector Machine creates a hyperplane in an n-dimension that classifies the data points. The hyperplane selected must have the maximum margin or the maximum distance between two data points of both classes [7]. Hyperplanes are decision boundaries that classify the data points, and the dimension of the hyperplane depends upon the number of features. An added benefit of SVM is not only does it work well with linearly separable data, but you can use various non-linear kernels to fit your data.

*Hyperparameters*

- Kernel specifies which kernel should be used. Usable kernels are linear, poly, and RBF. Linear uses linear algebra to solve for the hyperplane, while poly uses a polynomial to solve for the hyperplane in a higher dimension.
- Regularization is the coefficient used in the penalty function.

The hyperparameters were tuned to find the best ROC AUC score for the support vector machine using an exhaustive grid search (10-fold cv). The optimal hyperparameters are listed below.

- Kernel: Linear
- Regularization: C = 0.009

K-Nearest Neighbors Classifier

The K-Nearest-Neighbors algorithm calculates the distance of a new data point to all other training data points [8]. The distance metric can be of any type, such as Euclidean or Manhattan. Then, selects the K-nearest data points, where K can be any integer. Finally, it assigns the data point to the class to which the majority of the K data points belong.

*Hyperparameters*

- n_neighbors represents the number of neighbors for classification.
- leaf_size determines how many observations are captured in each leaf.
- metric can be set to various distance metrics like Manhattan, Euclidean, Minkowski, or weighted Minkowski.

The hyperparameters were tuned to find the best ROC AUC score for the KNN classifier using an exhaustive grid search (10-fold cv). The optimal hyperparameters are listed below.

- n_neighbors: 24
- leaf_size: 2
- metric (p): 1

Random Forest

Random forest is an ensemble method, which uses a combination of tree predictors such that each tree depends on the values of a random vector sampled independently and with the same distribution for all trees in the forest. The generalization error for forests converges as to a limit as the number of trees in the forest becomes large. The generalization error of a forest of tree classifiers depends on the strength of the individual trees in the forest and the correlation between them [9].

*Hyperparameters*

- n_estimators is the number of decision trees used in making the forest.
- criterion measures the quality of the split and receives either "gini", for Gini impurity (default), or "entropy", for information gain. Gini impurity is the probability of incorrectly classifying a randomly chosen data point if it were labeled according to the class distribution of the dataset. Entropy is a measure of chaos in your data set. If a split in the dataset results in lower entropy, then you have gained information.
- max_features is the number of features to consider when looking for the best split.
- max_depth is the maximum depth of the tree.
- min_samples_split is the minimum number of samples required to split an internal node:
- Bootstrap is a concept of repeatedly sample data with replacement from the original training set in order to produce multiple separate training sets. These are then used to allow ensemble methods to reduce the variance of their predictions, thus greatly improving their predictive performance.

The hyperparameters were tuned to find the best ROC AUC score for the KNN classifier using a randomized search (10-fold cv) [10]. The optimal hyperparameters are listed below.

- n_estimators: 300
- Criterion: entropy
- max_features: 4
- max_depth: None
- min_samples_split: 15
- Bootstrap: True

*Results*

The ROC curves with AUC scores are presented in Figure 13. More in-depth scoring was recorded in Table 3. The performance of each model is listed from best to worst, RF > KN > LR > SVM. The results are not a surprise as ensemble methods tend to do very well. And being the data is somewhat linearly separable, Logistic Regression and Support Vector Machine with a linear kernel would perform similarly. Overall, the ROC AUC scores are very promising, and the models have fit this dataset well. However, more investigation needs to be conducted to see if our models fit certain behaviors better (people who have higher average playing densities and average hours played).

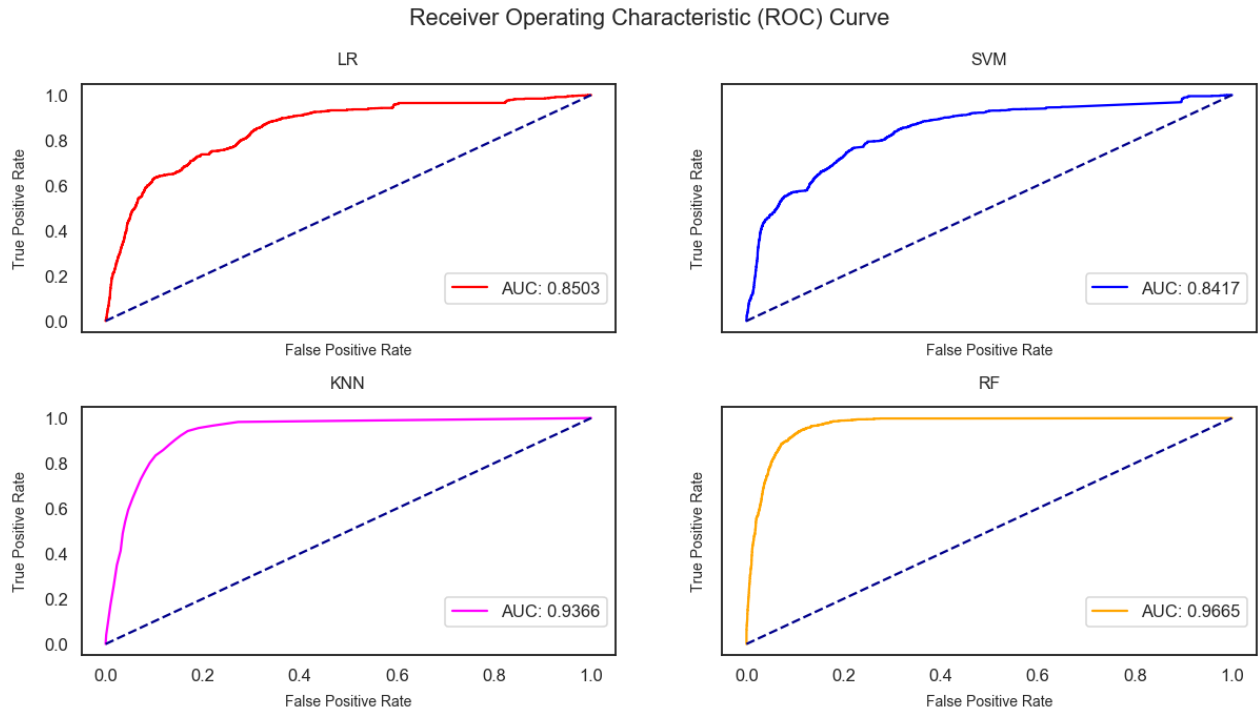

Figure 13: Receiver Operating Characteristic (ROC) curves with AUC scores for each machine learning classifier.

Table 3: Binary classification scores with various machine learning algorithms.

| Method | Accuracy | ROC_AUC | |
| --- | --- | --- | --- |
| | | 10-fold CV | Testing |
| Logistic Regression | 0.8451 | 0.8532 | 0.8503 |
| Support Vector Machine | 0.6384 | 0.8454 | 0.8417 |
| K-Nearest Neighbors | 0.8814 | 0.9407 | 0.9366 |
| Random Forest | 0.9171 | 0.9254 | 0.9665 |

*Alternative Techniques*

Along with the methods used above, three additional methods were used to aid in our binary classification problem: K–means clustering, PCA (2D) + K-means clustering, PCA (3D) + K-means clustering.

K-means clustering is an algorithm to classify or to group objects based on attributes or features into K number of groups [11]. The grouping is done by minimizing the sum of the square of distances between data and the corresponding cluster centroids. Generally, used as an unsupervised learning technique to find patterns amongst unstructured data or data without labels, K-means can be used for supervised learning problems such as classification. However, k-means does not compute probabilities in classifying labels. Therefore, the ROC AUC score cannot be computed, but the accuracy can be used instead.

Principal Component Analysis (PCA) is a dimensional reduction technique that preserves the essence of the original data. Having multi-dimensional data (many features), it is very important in identifying which features are important. PCA finds a new set of dimensions (or a set of basis of views) such that all the dimensions are orthogonal (and hence linearly independent) and ranked according to the variance of data along with them [12].

*Results*

The clustering plots of the training and testing sets are presented in Figure 14 (a, b, c, d, e, f). The 3D scatterplots of k-means clustering (a,b) and PCA (3D) + k-means clustering (e,f) look very similar.

(a) K-means clustering on training set.

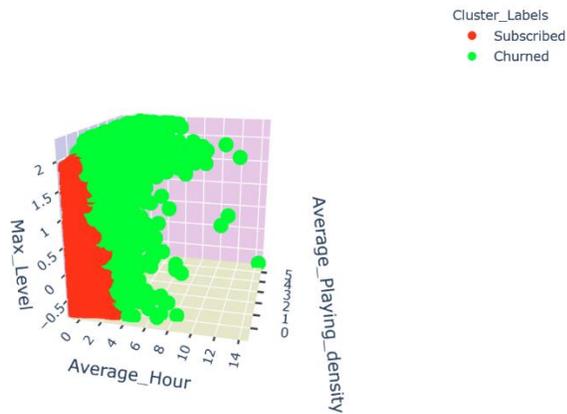

(b) K-means clustering on testing set.

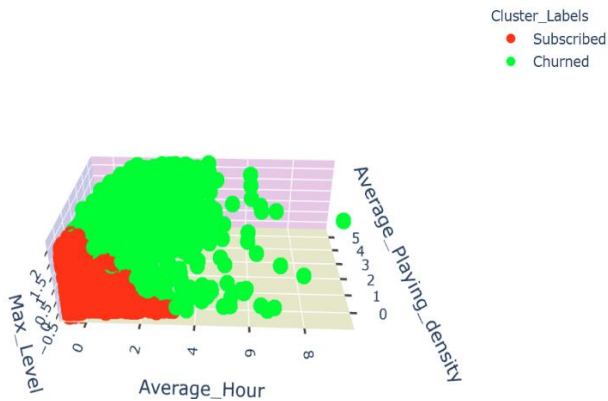

(c) PCA (2D) + K-means clustering on training set.

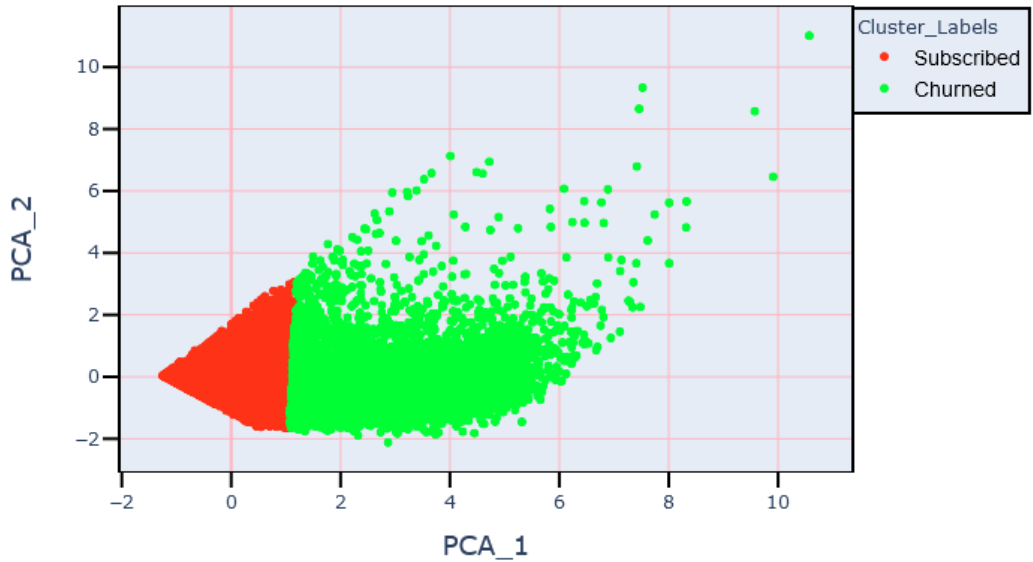

(d) PCA (2D) + K-means clustering on testing set.

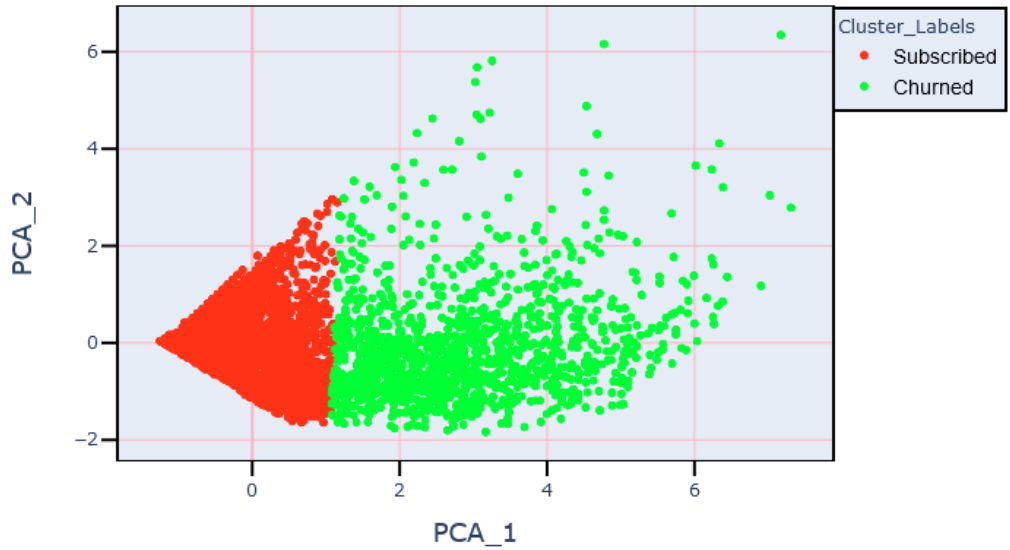

(e) PCA (3D) + K-means clustering on training set.

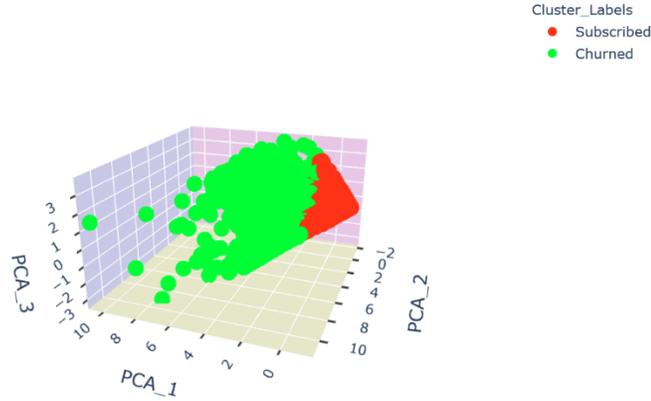

(f) PCA (3D) + K-means clustering on testing set.

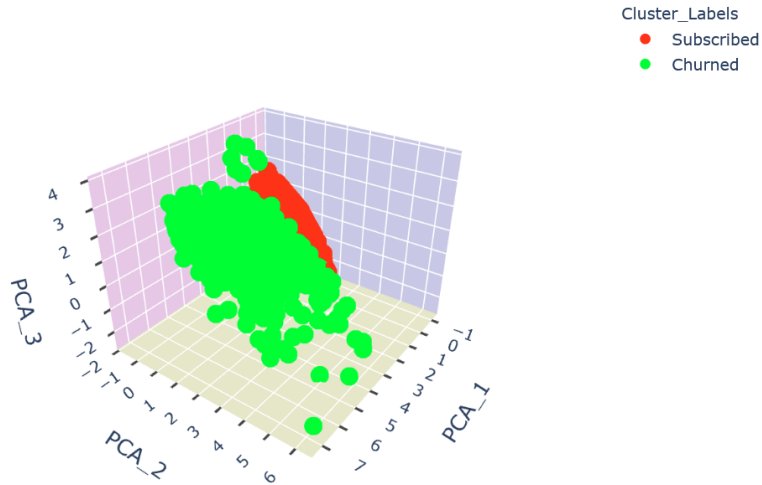

Figure 14: (a) K-means clustering on training set. (b) K-means clustering on testing set. (c) PCA (2D) + K-means clustering on training set. (d) PCA (2D) + K-means clustering on testing set. (e) PCA (3D) + K-means clustering on training set. (f) PCA (3D) + K-means clustering on testing set.

The number of PCA components is presented in Figure 15 (a, b, c). A general rule of thumb is to select the number of components that accounts for at least 85% or 95% of the variance, which are two or three components, respectively. The accuracies of the models are listed in Table 4. Overall, all methods performed similarly with an accuracy ~ 81% and is a good baseline for future models.

(a) PCA variance across multiple components.

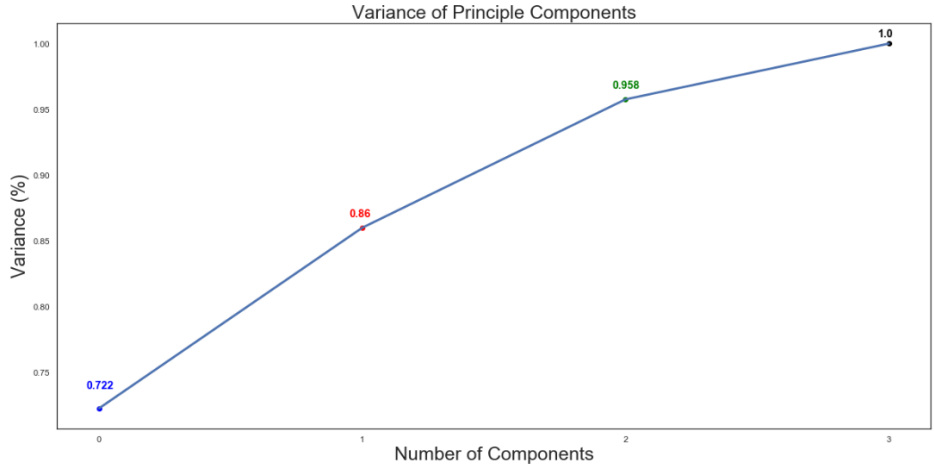

(b) Two Components.

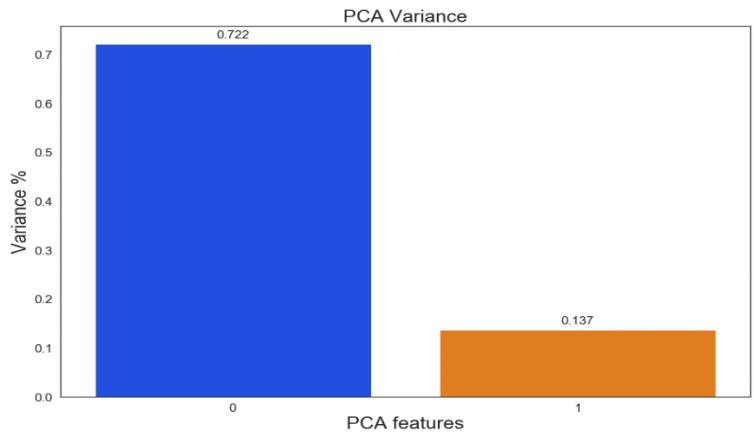

(b) Three Components.

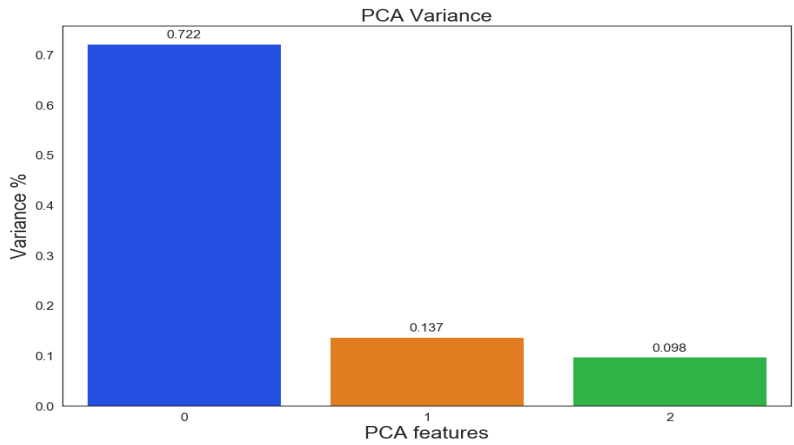

Figure 15: (a) PCA variance across multiple components. (b) Two components. (c) Three components.

Table 4: Binary classification scores with k-means clustering + PCA.

| Method | Accuracy | |
|---|---|---|
| | Training | Testing |
| K-Means Clustering | 0.817253 | 0.816758 |
| PCA 2D + K-Means Clustering | 0.817253 | 0.816891 |
| PCA 3D + K-Means Clustering | 0.817221 | 0.816891 |

## Conclusion

The survival analysis establishes a healthy financial model for Blizzard. With lifetimes of ~215 days until churn, it allows for Blizzard to make a hefty sum of money before a person will unsubscribe. There is still unpredictability due to how censored the data is – many entries have not churned in the one-year period and cannot judge past the 365-day duration. The survival curves for each feature exhibit expected behavior such as having higher average playing hours, average playing density or being in a guild results in a lower churn ratio. Lastly, the binary classification performed in the best performing algorithm having a 96% ROC AUC score in predicting whether a customer will churn in six months.

Possible future work includes

(1) Examine a three-year period to have more accurate analyses for survival analysis and six-month churn prediction.

(2) Study the performance of our algorithms with different playing patterns – does the data fit well for players who exhibit more playing time?